\documentclass[sn-mathphys,Numbered]{sn-jnl}

\usepackage{graphicx}%
\usepackage{multirow}%
\usepackage{amsmath,amssymb,amsfonts}%
\usepackage{amsthm}%
\usepackage{mathrsfs}%
\usepackage[title]{appendix}%
\usepackage{textcomp}%
\usepackage{manyfoot}%
\usepackage{booktabs}%
\usepackage{algorithm}%
\usepackage{algorithmicx}%
\usepackage{algpseudocode}%
\usepackage{listings}%

\usepackage{newtxtext,newtxmath}

\usepackage{physics}

\usepackage{graphicx}	
\usepackage{amsmath}	
\usepackage{amssymb}	
\usepackage{makeidx}

\usepackage[normalem]{ulem}
\usepackage{journal_abbreviations}

\usepackage{natbib}
\usepackage{cleveref}

\usepackage{enumitem}

\usepackage{wrapfig}
\usepackage{adjustbox}
\usepackage{subcaption}

\usepackage[dvipsnames]{xcolor}
\usepackage{tablefootnote}

\graphicspath{{./figures/}}

\usepackage{caption}

\usepackage{geometry}
\usepackage{pdflscape}
\usepackage{lscape}

\theoremstyle{thmstyleone}%
\theoremstyle{thmstyletwo}%

\theoremstyle{thmstylethree}%

\begin{document}

\title[Article Title]{
The miscibility of hydrogen and water in planetary atmospheres and interiors
}


\author*[1,2,3]{\fnm{Akash} \sur{Gupta}}\email{akashgpt@princeton.edu}

\author[3]{\fnm{Lars} \sur{Stixrude}}

\author[3]{\fnm{Hilke E.} \sur{Schlichting}}

\affil[1]{\orgdiv{Department of Astrophysical Sciences}, \orgname{Princeton University}, \orgaddress{\state{NJ}, \postcode{08544},  \country{USA}}}

\affil[2]{\orgdiv{Department of Geosciences}, \orgname{Princeton University}, \orgaddress{\state{NJ}, \postcode{08544}, \country{USA}}}

\affil[3]{\orgdiv{Department of Earth, Planetary, and Space Sciences}, \orgname{University of California, Los Angeles}, \orgaddress{\state{CA}, \postcode{90095}, \country{USA}}}


\abstract{

Many planets in the solar system and across the galaxy have hydrogen-rich atmospheres overlying more heavy element-rich interiors with which they interact for billions of years.  Atmosphere-interior interactions are thus crucial to understanding the formation and evolution of these bodies.  However, this understanding is still lacking in part because the relevant pressure-temperature conditions are extreme.  We conduct molecular dynamics simulations based on Density Functional Theory to investigate how hydrogen and water interact over a wide range of pressure and temperature, encompassing the interiors of Neptune-sized and smaller planets.  We determine the critical curve at which a single homogeneous phase exsolves into two separate, hydrogen-rich and water-rich phases, finding good agreement with existing experimental data.  We find that the temperature along the critical curve increases with increasing pressure and shows the influence of a change in fluid structure from molecular to atomic near 30 GPa and 3000 K, which may impact magnetic field generation. The internal temperatures of many exoplanets, including TOI-270 d and K2-18 b may lie entirely above the critical curve: the envelope is expected to consist of a single homogeneous hydrogen-water fluid, that is much less susceptible to atmospheric loss as compared with a pure hydrogen envelope. As planets cool, they cross the critical curve, leading to rainout of water-rich fluid and an increase in internal luminosity.  Compositions of the resulting outer, hydrogen-rich, and inner, water-rich envelopes depend on age and instellation and are governed by thermodynamics.  Rainout of water may be occurring in Uranus and Neptune at present.

}




\maketitle

H and O are the first and third most abundant elements in the universe, and hydrogen and water are major constituents of planets.  In our own solar system, the interiors of the ice giants, Uranus and Neptune, are thought to be primarily composed of water with an overlying hydrogen-rich envelope \citep{movshovitz2022a,scheibe2021a}. The cores of the giant planets, Jupiter and Saturn, may also be primarily composed primarily of water together with an uncertain admixture of rock \citep{guillot2015a,lainey2012a}, underlying a hydrogen-rich interior. {The mass and radius measurement of many exoplanets have also been interpreted as being consistent with a hydrogen-dominated atmosphere with water-rich interiors, i.e. water mass fractions of $\sim$30-50\% \citep{zeng2019a,madhusudhan2021a,luque2022a,holmberg2024a}.}

The thermal evolution of water-rich bodies has been widely studied, primarily in the context of strictly layered structures in which hydrogen-rich envelopes and water-rich interiors do not exchange mass \citep{nettelmann2013a,scheibe2019a}. However, the assumption of limited chemical interaction between hydrogen and water at planetary interior conditions may not be accurate{, and could have led to unrealistic inferences about the composition and structure of planet interiors and atmospheres}. Furthermore, a better understanding of the chemical reaction between hydrogen and water is important for the possible formation of dilute or fuzzy cores \citep{fuller2014a}, for the possible onset of rainout on cooling \citep{bailey2021a,stevenson1977b}, and for the dissolution of planetesimals in the H-rich envelope during late-stage accretion \citep{fortney2013a}.   

The nature and extent of chemical interaction between major planetary forming constituents is a major source of uncertainty in our understanding of planetary evolution.  The pressures and temperatures in planetary interiors often exceed current laboratory capabilities, and experimental constraints on the relevant phase equilibria are limited.  In the case of the hydrogen-water system, experimental observations of inter-solubility exist up to a pressure of 3 GPa, corresponding to $\sim$90\% of the radius of Uranus, leaving the vast majority of the interior of ice giants and Hycean worlds unconstrained by experiments \citep{seward1981a,bali2013a,vlasov2023a}.  The experimental data disagree with a previous theoretical study, which was not able to delineate the conditions under which hydrogen and water-rich fluids may coexist stably as two separate phases \citep{soubiran2015a}.

\begin{figure}
\centering

                \includegraphics[width=1.0\textwidth,trim=0 -100 0 0,clip]{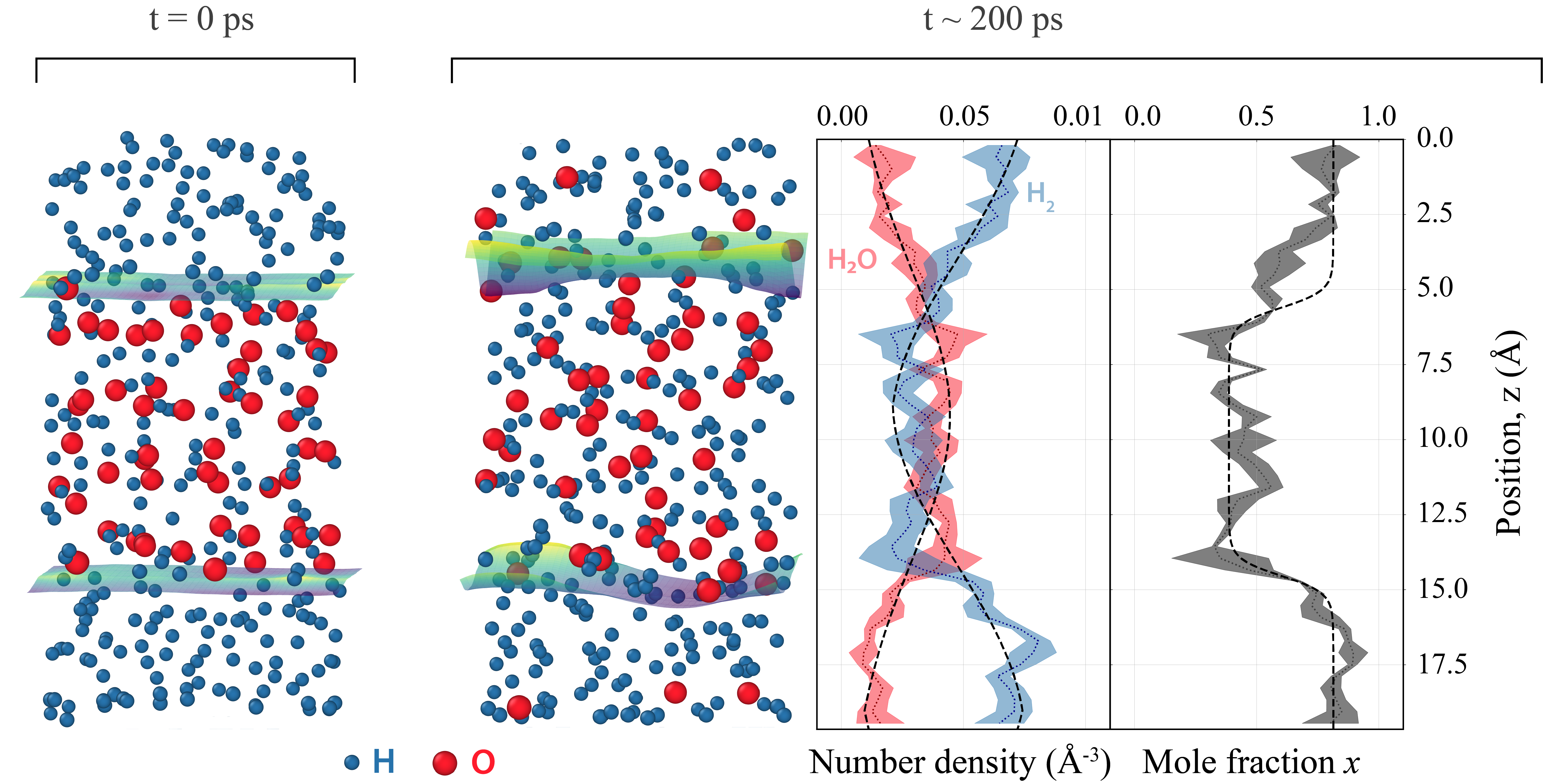}

    \caption{Simulation snapshots of the two-phase system at 1000 K and 10 GPa. Oxygen atoms are denoted by red spheres, and hydrogen atoms by blue spheres. The Gibbs dividing surface is illustrated by the green-blue surface, separating H-rich from H-poor phases \citep{willard2010a}. The initial condition is on the left and an equilibrated snapshot at 200 ps in the center. The rightmost panel shows the spatial variation of the H-concentration $x$ of the equilibrated portion of the simulation over a 5 ps interval, and the dashed line results from a fit to Eq. 2.
    }
    
     \label{fig:methodology}
\end{figure}

We explore the interaction between hydrogen and water over a wide range of pressures and temperatures that encompasses their complete miscibility, the molecular-to-atomic transition, and conditions expected during the evolution of planets.  We delineate the conditions under which hydrogen and water-rich fluids are completely miscible, determine the compositions of coexisting phases in the two-phase regime, explore the structure of the fluids to provide an atomistic level understanding of our results and examine the consequences of our results for the compositional, thermal and structural evolution of the envelopes and interiors of ice giants and Hycean worlds.

\section*{Results}\label{sec:results}

\begin{figure}
\centering
    \includegraphics[width=\textwidth,trim=100 0 0 0,clip]{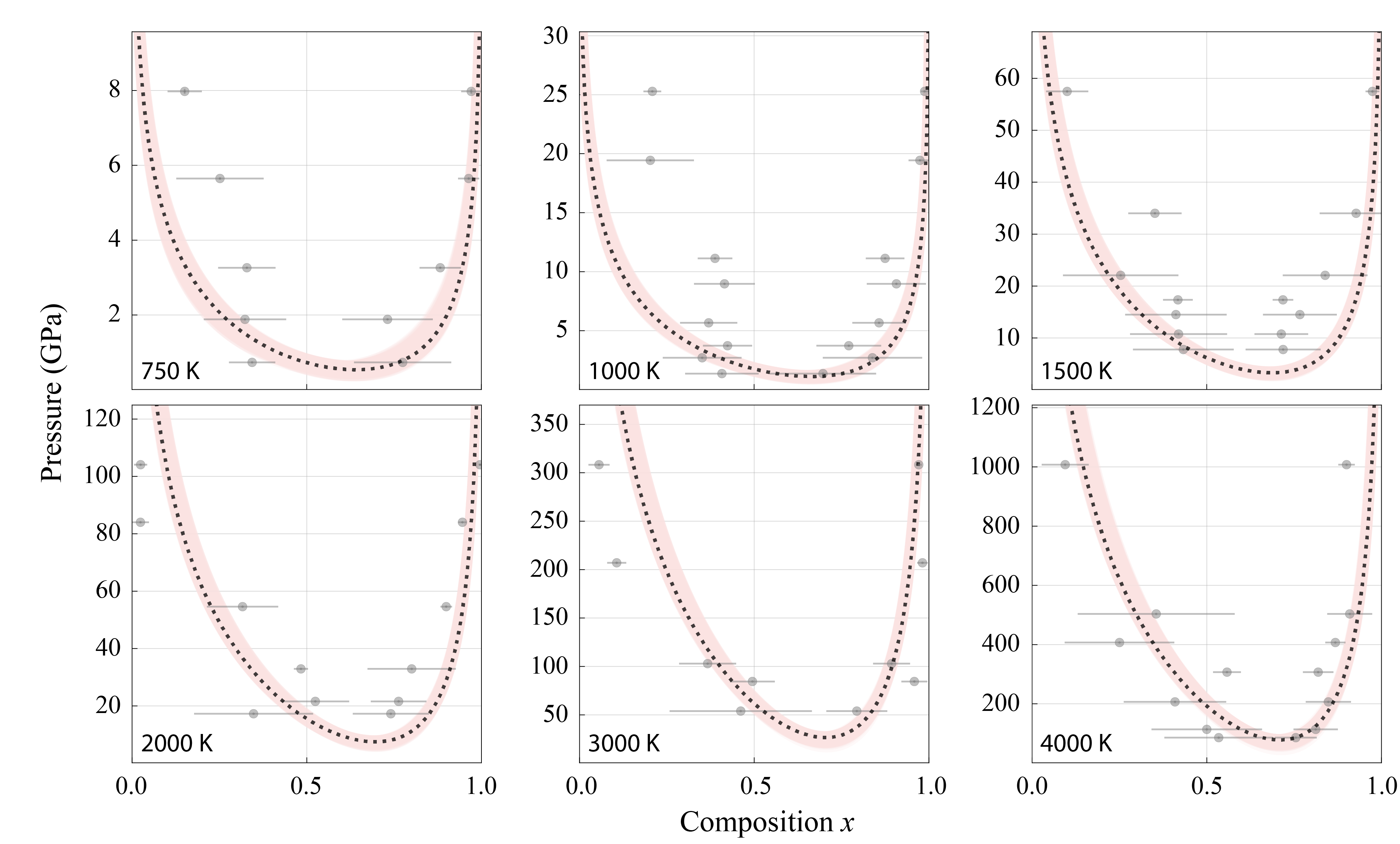}
    \caption{Coexistence curves from our simulations along several isotherms indicated, showing the compositions of the two coexisting phases as determined from our simulations (grey symbols with error bars), and best-fit coexistence curves to our simulation results (black dashed curves with pink uncertainty envelopes).
    } 
     \label{fig:coexistence_curves}
\end{figure}

\begin{figure}[ht!]
    \centering
        \includegraphics[width=\textwidth,trim=90 -50 15 40,clip]{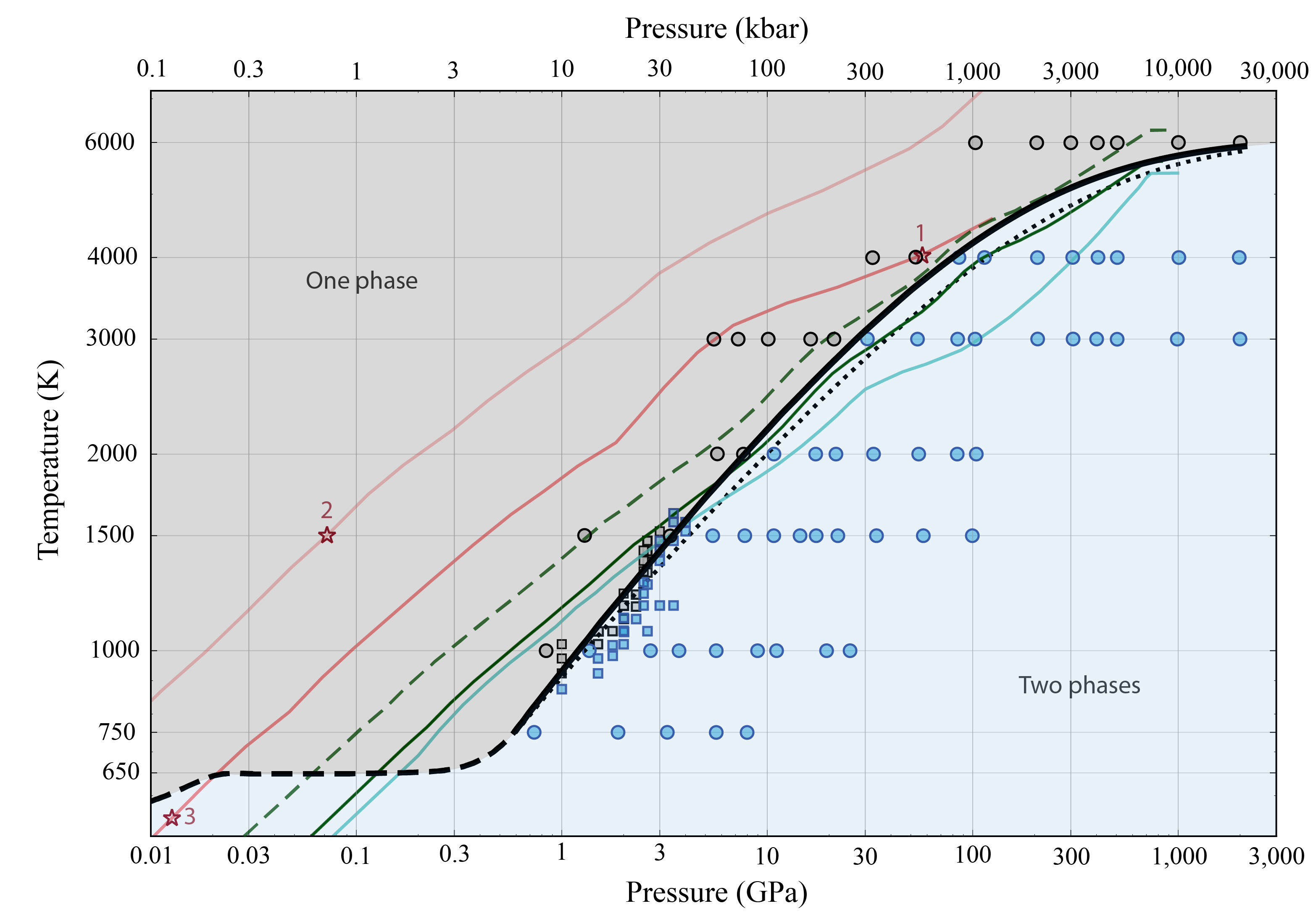}
        \caption{The pressure-temperature conditions of our simulations that result in two coexisting phases (blue circles) or a single homogeneous phase (gray circles), the critical curve derived from our results (bold black line) and as constrained by experiments at low pressure (bold dashed black line) \citep{seward1981a}, other experimental results (blue squares: two phases, gray squares: one phase) \citep{bali2013a,vlasov2023a}, the phase boundary computed from our results at $x=0.6$ (black dotted line), model temperature profiles for Uranus (green) and Neptune (blue) \citep{scheibe2019a}, and a hot Uranus temperature profile with $T_1$=85 K (dashed green).  The red lines represent possible loci of the boundary between the hydrogen-rich and the water-rich layer in models of the K2-18b according to \citet{madhusudhan2020a} for internal temperatures of 50 K (light red curve) and 25 K (dark red curve).  The numbered points (red stars) refer to the boundary between the hydrogen-rich and the water-rich layers in three possible internal structure models that are examined in more depth by \citet{madhusudhan2020a}.}
        \label{fig:critical_curve}
    \end{figure}

\subsection*{Phase Equilibria}

We perform two-phase simulations initiated as domains of pure water and pure hydrogen fluids joined at a planar interface (\Cref{fig:methodology}) \citep[]{xiao2018a}.  At high pressure, we find two separate phases, coexisting in dynamic equilibrium, each with stationary compositions: one relatively hydrogen-rich and the other hydrogen-poor (\Cref{fig:coexistence_curves}).  We quantify the compositions of the two coexisting fluids in terms of the mole fraction of the H$_2$ component
\begin{equation}
    x=\frac{N_{H_2}}{N_{H_2}+N_{H_2O}}=\frac{N_H-2N_O}{N_H}
\end{equation}
and $N_i$ are the number of molecules or atoms of type $i$.  At the highest pressure of our simulations, except at temperatures of 6000 K, the two coexisting fluids are nearly pure hydrogen and water, respectively.  With decreasing pressure, the compositions of the two phases approach each other.

Along each isotherm, we find a value of the pressure, the critical pressure, $P_c$, at which the two-phase compositions become identical.  At pressures less than the critical pressure, we find a single homogeneous phase in our simulations.  The critical pressure increases with increasing temperature (\Cref{fig:critical_curve}). The critical curve separates two stability fields: a higher temperature stability field in which only one homogeneous phase is stable, and a lower temperature stability field in which two separate phases may exist. 

Our results are in excellent agreement with available experimental data \cite{bali2013a,vlasov2023a} (\Cref{fig:critical_curve}).  While we are not able to probe the phase diagram at temperatures lower than 750 K, our results are consistent with experiments in this low-temperature limit \cite{seward1981a}. Our results are in better agreement with the experiment than a previous computational study, which relied on a parameterized force field \cite{bergermann2021a}, instead of density functional theory, as we have done. A previous study based on density functional theory, but using a methodology based on the computation of free energies rather than phase coexistence, inferred significantly higher critical temperatures as compared with our results or those of previous experiments \cite{soubiran2015a}.    

Our results show that the critical temperature increases with increasing pressure, at least up to 2000 GPa (\Cref{fig:critical_curve}).  The slope of the critical curve $\nabla_c=d \ln T_c/d \ln P$ is nearly constant over the lower pressure range of our simulations (1-30 GPa), where it has a value $\nabla_c=0.36$.  At pressure greater than 30 GPa, $\nabla_c$ decreases markedly with increasing pressure and becomes almost independent of pressure at the highest pressure limit of our simulations.  

\begin{figure}[ht!]
    \centering
        \includegraphics[width=\textwidth,trim=0 -20 0 0,clip]{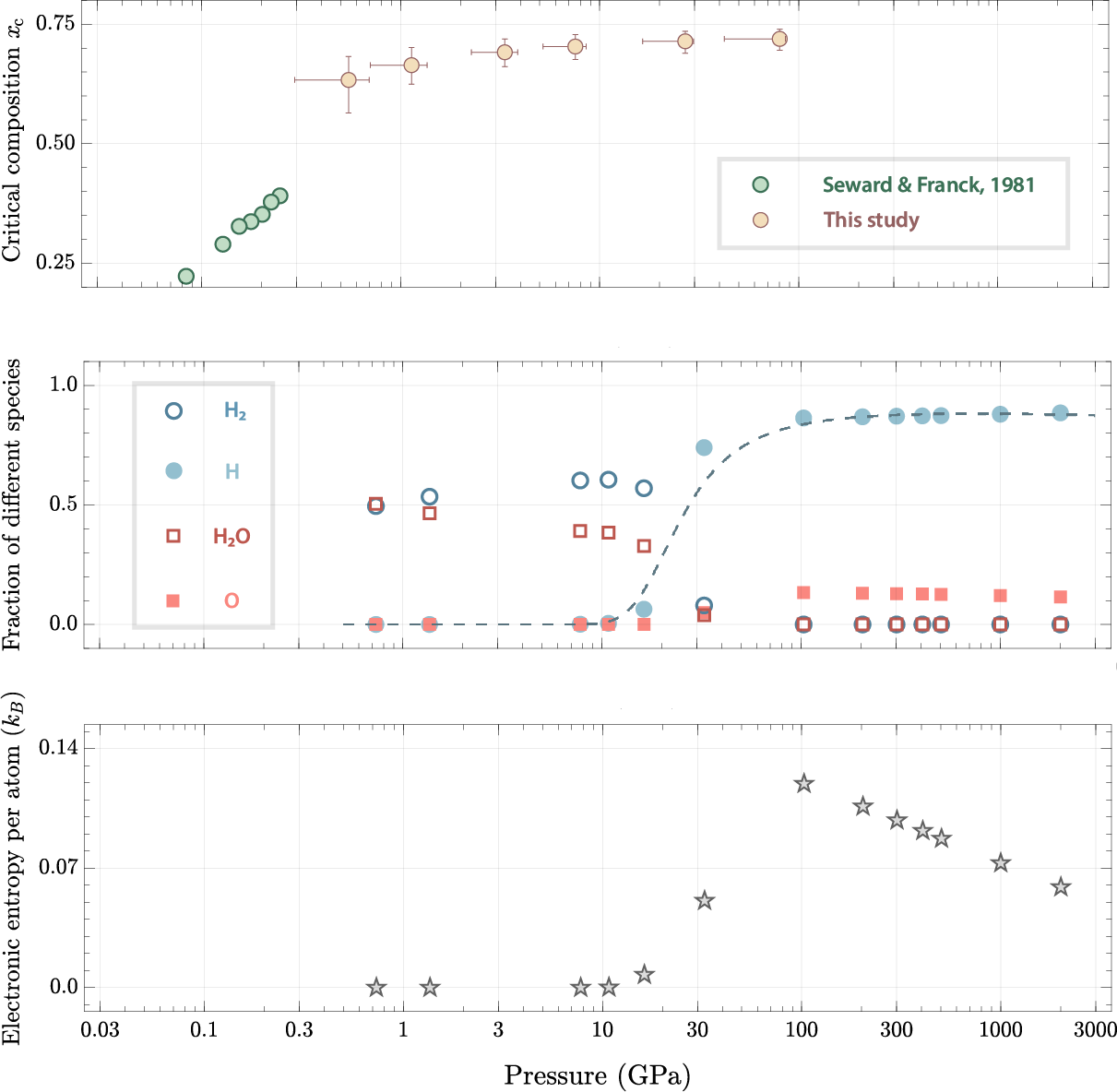}
        \caption{Variation of the critical composition of the hydrogen-water system with pressure (top), speciation (middle) and electronic entropy (bottom) of the homogeneous fluid along the critical curve. In the top panel, the green dots show the experimentally determined values of the critical composition and its comparison with those from this study. The middle panel shows the fraction of the species indicated in the homogeneous fluid along the critical line. The blue dashed curve is based on a Fermi-Dirac-like function tracing the change in the H-fraction and showcases the molecular to atomic transition in the fluid at $\sim$30-100 GPa and 3000-4000 K. The bottom panel shows the change in the electronic entropy along the critical curve and indicates the metallization of the homogeneous fluid as it transitions from molecular to atomic.
        }
        \label{fig:multiple_props}
    \end{figure}

The critical curve delineates the boundary between one-phase and two-phase stability at the critical composition $x_c$ (\Cref{fig:multiple_props}).  The critical composition increases with increasing pressure from 0.63 to 0.71 over the range of our simulations.  The value of $x_c$ that we find, therefore, lies on the H$_2$-rich side of the H$_2$-H$_2$O join ($x_c$$>$0.5).  The increase of $x_c$ with increasing pressure that we find is in excellent agreement with experiments at lower pressures, which find that $x_c$ increases towards 0.4 at the highest experimental pressure (0.3 GPa).  

Our results also allow us to determine the conditions of two-phase stability at bulk compositions other than the critical composition.  For example, the ice giants are expected to be more water-rich in bulk composition than the value of $x_c$ that we find.  For any composition differing from the critical composition, the boundary between one-phase and two-phase stability is shifted to higher pressure (lower temperature) as compared with the critical curve.  For example, at 2000 K, the phase boundary increases from 7 GPa at the critical composition to 15 GPa at a bulk composition representative of the ice giants: $x$=0.6, corresponding to a bulk heavy element mass fraction $Z$=0.85 \citep{scheibe2019a}.

\subsection*{Structure and Bonding}

We find that the change in the slope of the critical curve is caused by a change in the structure of the fluid (\Cref{fig:multiple_props}).  The fluid on the critical curve undergoes a molecular-to-atomic transition near 30 GPa and 3000 K: dominated by H$_2$O and H$_2$ molecules at lower pressure, and H and O atoms at higher pressure.  We find no evidence of H$_3$O molecules in our simulations.  It had been proposed that H$_3$O may be an important constituent in the deep interiors of the ice giants \citep{huang2020a}.  This proposal was based on the analysis of crystalline structures.  Evidently, the structure of the fluid differs fundamentally from that of crystals at the same pressure, and H$_3$O becomes unstable to dissociation at high temperatures.

Along with the change in speciation, the electronic structure of the fluid also changes, from insulating to metallic: the electronic entropy is zero at low pressure when the fluid is molecular, but increases rapidly as the fluid becomes atomic (\Cref{fig:multiple_props}). The conditions of the molecular to atomic and metallization transitions that we find along the critical curve are similar to those of the insulating to metallic transition in pure hydrogen as seen experimentally \cite{mcwilliams2016a}.

The molecular-to-atomic transition may have important implications for our understanding of the magnetic fields of water-hydrogen planets (\Cref{fig:multiple_props}).  In the atomic regime, the electrical conductivity is much greater than it is in the molecular regime, where H is the dominant charge carrier.

\section*{Discussion}

We expect the phase transition that occurs on the crossing of the critical curve to be an important part of the compositional, thermal, and structural evolution of water-hydrogen planets.  The reason is that the temperature and the temperature gradient of the critical curve are very similar to those expected in the interior of many planets, including the ice giants (Fig.~\ref{fig:critical_curve}).

As a water-hydrogen planet cools, the temperature profile crosses the critical curve, hydrogen-rich and water-rich phases separate, and the denser water-rich fluid rains out, producing a layered structure in which a hydrogen-rich envelope surrounds a water-rich interior.  Such a layered structure is consistent with our knowledge of the interiors of the ice giants.  For instance, the pressure at which the crossing first occurs, near 30 GPa, is similar to the pressure at which the boundary between hydrogen-rich and water-rich layers are inferred in models of the interiors of Uranus and Neptune that are consistent with observational data such as the gravity moments \citep{nettelmann2013a,scheibe2019a,movshovitz2022a}.  Interior models of the ice giants are, however, uncertain, and the pressure at which the planet first crosses the critical curve depends on several factors, including its bulk composition.  For example, for a bulk composition that differs from the critical composition and which is more typical of the ice giants ($x$=0.6), the critical curve shifts to higher pressure by 10 GPa at 3000 K.

As the planet continues to cool, the crossing shallows, producing further rainout of water, and growth of the water-rich interior at the expense of the hydrogen-rich envelope.  The gravitational energy released by the ongoing concentration of the denser component at depth provides an energy source that can significantly influence the planet's thermal evolution.  The idea that phase separation influences a planet's luminosity was first proposed in the context of the gas giants and the separation of He from H \citep{stevenson1977a,stevenson1977b}.  The possibility of phase separation of water from hydrogen in the ice giants has also been explored but on the basis of a critical curve and critical composition derived from extrapolation of limited experimental data, which differs substantially from ours \citep{bailey2021a}.  In comparison with their critical curve, ours is much shallower at high pressure ($\nabla_c$ is smaller), producing phase separation that is much deeper and later in a planet's evolution.  For example, within present uncertainties on the temperature profile, it is possible that Uranus has not yet reached the critical curve and that Neptune has (\Cref{fig:critical_curve}), which may contribute to the greater luminosity of Neptune as compared with Uranus.  Including the effects of phase separation at the critical curve in thermal evolution models of hydrogen-water planets will be important for understanding these bodies.  

As a planet cools and undergoes phase separation, the envelope and interior layers are not pure, but have compositions that evolve and are dictated by the phase equilibria (\Cref{fig:coexistence_curves}). This means, for example, that the hydrogen-rich envelope is expected to contain water, and the composition or metallicity of the envelope depends on a planet's age and the instellation it receives. Evolutionary scenarios may, therefore, be compared with the amounts of water spectroscopically detected in hydrogen-rich envelopes.

Our results have implications for our understanding of the origin of the ice giants.  Although the details of the interior structure of the ice giants are poorly known, substantial variation of composition with depth is required by their mean density, which far exceeds that of a planet made entirely of the material in the visible, hydrogen-rich envelope \citep{movshovitz2022a}.  The origin of the layered structure may be seen as the result of the accretion of an ice-rich core and subsequent capture of a hydrogen-rich envelope \citep{helled2020a}.  However, our results show that such a structure is thermodynamically unstable in a young, hot planet that lies above the critical curve: the water-rich and hydrogen-rich layers dissolve in each other, tending towards a homogeneous composition.  Reaction reduces the size of the originally accreted core and may dissolve it entirely.  Moreover, any late accreted ice dissolves in the envelope.  In this view, the layering that we see today is the result not of accretion, but of much later phase separation as these planets cooled to the point that they encountered the critical curve.  An important uncertainty is the rate at which an initially layered and hot young planet is able to re-homogenize, a process that likely involves double-diffusive or turbulently diffusive convection \citep{leconte2012a}.  

{The vast majority of exoplanets discovered so far are considerably hotter than Uranus and Neptune \citep{bean2021a} and have temperature profiles that lie above the critical curve.  The majority of water-rich exoplanets planets, therefore, do not have a separate hydrogen-rich envelope, as it dissolves in the interior.  A one-phase water-hydrogen envelope has a higher mean molecular weight than that of a pure hydrogen atmosphere, similar to recent observations of the atmosphere of TOI-270 d \citep{benneke2024a}. Furthermore, given the high mean molecular weight, such an envelope is also less susceptible to atmospheric loss. However, in the radiative part of the atmosphere above the homopause, diffusion could lead to increasing hydrogen abundance with increasing height.  Nevertheless, the prevalence of planets with such high mean-molecular weight envelopes could potentially explain why atmospheric outflows have not been detected in H$\alpha$ and He 1083 nm spectroscopic lines from planets that are otherwise expected to be losing their H atmospheres at rapid rates \citep{dossantos2023a}. }

An exception to this scenario are the Hycean worlds in which the hydrogen envelope is so thin that the boundary between the water-rich interior and the hydrogen envelope occurs at very low pressure (1 bar).  For example, models of K2-18b explore a layered structure including a hydrogen-rich envelope overlying a water-rich layer \citep{madhusudhan2020a}.  The possible loci of the boundary between these two layers lie mostly above the critical curve and so are not in equilibrium (\Cref{fig:critical_curve}).  Of those examined by \cite{madhusudhan2020a} in more detail, only case 3 is consistent with the presence of an equilibrium boundary between a water-rich and a hydrogen-rich layer. {As another example, TOI-270 d, if dominated by hydrogen and water and assuming that its temperature at 1 bar is at least as large as its equilibrium temperature (387 K), does not have a separate water ocean, but instead a one-phase water-hydrogen envelope \citep{benneke2024a,holmberg2024a}.}

\section*{Conclusion}\label{sec:conclusion}

The chemical reaction between hydrogen and water is likely to affect the evolution of a wide range of planets, including the ice giants in our solar system and water-rich exoplanets. We find that in planets that are hotter than Uranus and Neptune, and in the ice giants early in their evolution, hydrogen and water are completely miscible over the entire range of pressure-temperature conditions in the envelope and interior.  Primary accreted ice-rich cores are thermodynamically unstable and react with an overlying H-rich envelope, limiting the size of the core and altering its density and structure.  Late-accreted planetesimals are likely to dissolve completely on their passage through the H-rich envelope, provided efficient fragmentation \citep{fortney2013a}, increasing the heavy element concentration of the envelope.  For planets as cold or colder than Uranus and Neptune, phase separation and rainout of H$_2$O occur, providing a source of gravitational energy that alters the planet's thermal evolution.

Our results also provide a framework for understanding the evolution of the composition of the envelope and making predictions of elemental abundances that may be detected by future missions and surveys. For example, the proposed Uranus Orbiter and Probe mission would measure the heavy element concentrations, including O, in the outer envelope of Uranus, which is influenced by the phase equilibria that we predict and rainout of H$_2$O.  The James Webb Space Telescope will measure the elemental abundances in the outer envelopes of hot exoplanets where H and H$_2$O are completely miscible, whereas the next generation direct imaging surveys will be able to probe the compositions of Neptune-like planets across a range of ages and equilibrium temperatures, providing important constraints on formation and evolution processes.

\newpage




\section*{Methods}\label{sec:methodology}

\subsection*{Computation}\label{sec:setup}

To study the hydrogen-water system, we employ Born-Oppenheimer \textit{ab-initio} molecular dynamics based on the Density Functional Theory \citep{kohn1965a}. We use the projector-augmented wave method as implemented in the Vienna \textit{ab-initio} Simulation Package \citep[VASP;][]{kresse1993a,kresse1996a,kresse1996b,kresse1999a}. H and O have valence configurations of (1s$^1$) and  (2s$^2$ 2p$^4$), and core radii of 1.100 and 1.520 Bohr, respectively. For the exchange-correlation potential, we use the PBEsol approximation \citep{perdew2008a}. Previous studies have demonstrated that PBEsol yields good agreement with experiment \citep{scipioni2017a,holmstrom2018a}. We sample the Brillouin zone at the Gamma point and use a plane-wave cutoff of 500 eV, {which we find yield pressure and energy convergence to within 0.78 GPa and 2 meV/atom, respectively.} We assume thermal equilibrium between ions and electrons via the Mermin functional \citep{mermin1965a,wentzcovitch1992a}.

We performed canonical ensemble (\textit{NVT}) simulations using the Nos\'e-Hoover thermostat \citep{hoover1985a}. We first perform homogeneous one-phase simulations of pure hydrogen and pure water.  The number of water molecules (54) is the same in all simulations with the volume of the cell $V$ chosen to yield the desired pressure.  Hydrogen simulations have the same cell volume $v$, and the number of H atoms are adjusted to attain the desired pressure.  These pure phase simulations are run with a time-step of 0.5 or 1 fs chosen so that the drift in the conserved quantity is less than 2 meV/atom/ps. Depending on the $\{ T$, $P \}$ conditions, the equilibration of pure phases requires 5 - 50 ps. 

We then initiate the two-phase run by combining equilibrated homogeneous phase configurations as shown in the leftmost panel of \Cref{fig:methodology}. The total number of atoms in the two-phase simulations are 300-450, depending on the temperature and pressure. We run these simulations until the phase compositions are stationary, which typically requires 150-300 ps.

\subsection*{Analysis of simulations}\label{sec:results:analysis}

\textbf{Phase compositions}.  To determine the composition of these coexisting phases, we first estimate the number densities of the two species $\rho_H(z)$, $\rho_O(z)$, by averaging over 10,000 simulation steps; we estimate uncertainties via the blocking method  \citep{flyvbjerg1989a}. The resulting mole fraction of hydrogen $X (z) = \rho_H(z)/{(\rho_H(z) + \rho_O(z))}$ is then fit to a theoretically motivated hyperbolic tangent profile \citep{cahn1958a,widom1982a} 
\begin{align}
    X (z) &= X_2 + \frac{X_1 - X_2}{2} \sum_{j=1,2} (-1)^{j}
 \text{tanh}\left( \frac{(z -z_1) - \text{nint}(z - z_1) + (-1)^{j} w)}{\delta} \right).
\end{align}
where $X_1$ is the H fraction in the phase whose center of mass is located at $z_1$ and $X_2$ is the H fraction in the other phase, $w$ is the half-width of the phase, $\delta$ is the width of the interface; the nint function accounts for periodic boundary conditions, and the sum accounts for the presence of two interfaces. 

\textbf{Speciation}.  We determine the fraction of different species in equilibrated fluids from our simulations just above the critical curve, including molecules, and free atoms.  We search for a variety of species, not all of which turn out to be present in significant quantities, including H, O, H$_2$, O$_2$, OH, H$_2$O, and H$_3$O. We define that a set of atoms are bonded if they are in mutual proximity for some minimum time interval $\tau$ \cite[e.g.,][]{tamblyn2010a}, with $\tau$ chosen to be the period of the water bending mode (200 fs) \citep{seki2020a}. We define the neighborhood of atom A to consist of its five nearest neighbors.  Any atom or set of atoms that are within the neighborhood of A for the entire duration $\tau$ are considered to be bonded to A.

\subsection*{Thermodynamics}

We use a theoretically motivated model to describe our coexistence curves across the entire 
temperature pressure space of our simulations. The Gibbs free energy of mixing
\begin{equation}
    G=RT\left[ y \ln y + (1-y) \ln (1-y) \right] + Wy(1-y)   
\end{equation}
with, respectively, ideal and excess contributions on the right-hand side.  The compositional variable
\begin{equation}
y=\frac{x}{x+\lambda (1-x)}
\end{equation}
where $\lambda$ accounts for the asymmetric shape of the coexistence curves (Fig.~\ref{fig:coexistence_curves}).  We allow the excess Gibbs free energy parameter $W$ to depend on pressure and temperature
\begin{equation}
    W = W_H - T W_S + P W_V
\end{equation}
where $W_H$, $W_S$, and $W_V$ account for the excess enthalpy, entropy, and volume of solution, respectively \citep[e.g.,][]{haselton1980a,thompson1967a,dekoker2013a,saxena2012a}.  We assume that $W_H$ and $W_S$ are constants, and that
\begin{align}
W_V &= W_{V,1} + \frac{W_{V,2}}{(T/T_0)^2} \label{eq:W_V_eff} \\
\lambda &= \lambda_{1} + \frac{\lambda_{2}}{(T/T_0)}. \label{eq:lambda_eff}
\end{align}
where $T_0 = 1000$ K is a constant.  Finding the root $dG/dy=0$ and re-arranging
\begin{equation}\label{eq:coexistence_curve_P}
        P  = \frac{1}{W_V} \left[ \frac{RT}{(1-2y)} \ln \frac{1-y}{y} - \left( W_H - T W_S \right) \right]
\end{equation}
which yields the coexistence curves with the compositions of the two coexisting phases at $y$ and $1-y$.  We find the critical pressure by evaluating Eq.~\ref{eq:coexistence_curve_P} in the limit $y \rightarrow 1/2$
\begin{equation}
        P_c=\frac{2RT-(W_H-TW_S)}{W_V}
\end{equation}
and the critical composition
\begin{equation}
    x_c=\frac{\lambda}{1+\lambda}.
\end{equation}
We determine the most likely set of parameters $\Theta \in \{ W_{H}, W_{V,1}, W_{V,2}, W_{S}, \lambda_{1}, \lambda_{2} \}$ by fitting our simulation results (Fig.~\ref{fig:coexistence_curves}) to Eq.~\ref{eq:coexistence_curve_P}, using Bayesian inference with the Dynamic Nested Sampling method \citep{skilling2004a,skilling2006a,higson2019a} as implemented in the open-source code \texttt{dynesty} \citep{speagle2020a,koposov2023a}.

\begin{table}
    \centering
    {
    \renewcommand{\arraystretch}{2}
    \begin{tabular}{c  c  c  c} 
        \hline
            Parameter &  \multicolumn{3}{c}{ Values }\\
            & $\mu_{1/2}$ (50$^{th}$ percentile) & [$\mu_{1/2}-\sigma$, $\mu_{1/2}+\sigma$] & Units\\
        \hline
            $W_H$ & -599.08 & [-751.08, -401.18] & J mol$^{-1}$\\
            $W_{V,1}$ & -26.12 & [-27.46, -24.90] & J mol$^{-1}$ GPa$^{-1}$\\
            $W_{V,2}$ & 981.78 & [941.68, 1024.82] & J mol$^{-1}$ GPa$^{-1}$ K$^2$\\
            $W_S$ & -16.08 & [-16.20, -16.00] & J mol$^{-1}$ K$^{-1}$\\
            $\lambda_{1}$ & 2.62 & [2.56, 2.67] & - \\
            $\lambda_{2}$ & -0.68 & [-0.80, -0.62] & - \\
        \hline
    \end{tabular}
    \caption{Estimated parameters for coexistence curves across temperature-pressure space explored in this study $\in$ [750, 6000] K and [0.25, 2000] GPa.}
    \label{tab:coexistence_curve_parameters}
    }
\end{table}



\bibliography{planet_evo}

\section*{Acknowledgements}
A.G. thanks Leslie Insixiengmay and Adam Burrows for insightful discussions and is grateful for the support from the National Aeronautics and Space Administration (NASA), the Heising-Simons Foundation and Princeton University for the grant Future Investigators in NASA Earth and Space Science and Technology (FINESST; 80NSSC20K1372), the 51 Pegasi b Fellowship, and the Harry H. Hess Fellowship and Future Faculty in Physical Sciences Fellowship, respectively. A.G. also acknowledges the support from the Center
for Matter at Atomic Pressures (CMAP), a National Science Foundation (NSF) Physics Frontier Center, under Award PHY-2020249.  In addition, L.S. acknowledges support from the NSF under grant EAR-2223935 and H.E.S. acknowledges support from NASA under grant 80NSSC21K0392 issued through the Exoplanet Research Program.

This work used computational and storage services associated with the Hoffman2 Shared Cluster provided by UCLA Institute for Digital Research and Education's Research Technology Group, and Princeton Research Computing resources at Princeton University which is a consortium of groups led by the Princeton Institute for Computational Science and Engineering (PICSciE) and Office of Information Technology's Research Computing.

\section*{Author Contributions}
A.G. performed the calculations and data analysis. L.S. led the project design. All authors contributed to interpreting the data and writing the manuscript.

\section*{Competing Interests}
The authors declare no competing interests.

\section*{Materials \& Correspondence}
All correspondence should be addressed to A.G.

\end{document}